\documentclass[prd,nofootinbib,showpacs]{revtex4}
\usepackage{graphics}
\usepackage{bm}

\def\be{\begin{equation}}
\def\ee{\end{equation}}
\def\ba{\begin{eqnarray}}
\def\ea{\end{eqnarray}}
\def\dper{d_\perp}
\def\Dp{{\rm D}p}
\def\rp{r_\perp}
\def\rpl{r_\parallel}

\def\Vpl{V_\parallel}
\def\lp{\ell_\perp}
\def\lpl{\ell_\parallel}

\def\vpl{v_\parallel}

\def\linf{L_{\rm inf}}
\def\Deltar{\Delta^2_{\cal R}(k)}
\def\Deltarz{\Delta^2_{\cal R}(k_0)}
\begin{document}

\title{\large \bf Observational Constraints on Cosmic String Production
During Brane Inflation}
\author{Levon Pogosian$^1$, S.-H.~Henry Tye$^2$, Ira Wasserman$^{2,3}$
and Mark Wyman$^{2,3}$ \vspace{0.2cm}}
\affiliation{$^1$ Theoretical Physics, The Blackett Laboratory, Imperial
College, London, SW7 2BZ, United Kingdom \\
$^2$ Laboratory for Elementary Particle Physics, Cornell University,
Ithaca, NY 14853, USA \\
$^3$ Center for Radiophysics and Space Research, Cornell University,
Ithaca, NY 14853, USA}
\date{today}


\begin{abstract}
Overall, brane inflation is compatible with the recent analysis
of the WMAP data. Here we explore the constraints of WMAP and 2dFGRS data
on the various brane inflationary scenarios. Brane inflation
naturally ends with the production of cosmic strings, which may provide
a way to distinguish these models observationally. We argue that
currently available data cannot exclude a non-negligible contribution
from cosmic strings definitively. We perform a partial statistical analysis
of mixed models that include a sub-dominant contribution from cosmic
strings. Although the data favor models without cosmic strings,
we conclude that they cannot definitively rule out a 
cosmic-string-induced contribution of
$\sim 10 \%$ to the observed temperature, polarization and galaxy density
fluctuations. These results imply that $G\mu\lesssim
3.5\times 10^{-7}(\lambda/0.25)\sqrt{B/0.1}$, where $\lambda$ is a dimensionless 
parameter related to the interstring distance,
and $B$ measures the importance of perturbations induced by cosmic strings.
We argue that, conservatively, the data available currently still permit
$B\lesssim 0.1$.
Precision measurements sensitive to the B-mode polarization produced by
vector density perturbation modes driven by the string network could
provide evidence for these
models. Accurate determinations of $n_s(k)$, the scalar fluctuation index,
could also distinguish among various brane inflation models.
\end{abstract}

\pacs{98.80.Cq}

\

\maketitle

\section{Introduction}

Observations of the cosmic microwave background (CMB) \cite{cobe,new}
support the idea that the standard big bang phase of the expansion of
the Universe was preceded by inflation \cite{guth}. Recent results
from the Wilkinson Microwave Anisotropy Probe (WMAP) 
\cite{wmap_bennett,wmap_spergel,wmap_verde,wmap_peiris,wmap_komatsu}
constrain the properties of proposed inflationary models tightly,
but although some models are now excluded, numerous possibilities
remain. A further challenge to observational cosmology is to try to
hone in on a small class of viable models, even if identifying a
single, correct theory of inflation may prove impracticable.

All of the data collected up until now are consistent with a relatively
pristine Universe in which the perturbations observed today result
from the amplification and distortion of a relatively featureless,
Gaussian spectrum of fluctuations produced by quantum effects during
inflation. However, it is likely that inflation itself could have left
behind other remnants -- such as cosmic strings -- which could
actively perturb both the CMB and dark matter of the Universe up to the
present day.

It is well-known that cosmic strings cannot be wholly responsible for
either the CMB temperature fluctuations or the observed clustering
of galaxies \cite{aletal}; roughly speaking, the implied limits
on the cosmic string tension $\mu$ allowed by observations is 
$G\mu \lesssim 10^{-6}$.
However, now that cosmology has entered an era in which the properties
of the Universe are being revealed to unprecedented precision, a
natural question is to what extent the observations can allow previously
unwanted ingredients, such as cosmic strings (e.g. \cite{bouchet}).
Indeed, as the precision of cosmological observations increases, we might
hope to be able to distinguish among numerous presently viable models
for inflation by the properties of the cosmic strings they predict.

Although the idea that inflationary cosmology might lead to cosmic
string formation is not new \cite{kofman},
it has received new impetus from the
brane world scenario suggested by superstring theory. In brane
world cosmology, standard model particles and interactions correspond 
to open string (brane) modes, while the graviton, the dilaton and
the radions are closed string (bulk) modes. Thus, our 3D Universe can
be thought of as residing on a brane or stack of branes with three
dimensions of cosmological size. These branes in turn reside in extra 
dimensions that are compactified. In such a model, inflation can
result during the collisions of branes coalescing to form, ultimately,
the brane on which we live \cite{dvali-tye}.

In these brane inflation models, the
separations between branes in the compactified dimensions are
scalar fields (open string modes) that can act as inflatons, with 
the interaction
potential between spatially separated branes providing the
inflaton potential. Details of the brane inflation
scenario depend on both qualitative and quantitative features, such
as whether collisions involve a brane-antibrane pair \cite{burgess} 
or two branes coalescing at an angle \cite{rabadan}, as well as 
parameters such as the sizes of
the compactified dimensions \cite{collection,jst}. Qualitatively,
though, it appears easy to find models that predict adiabatic
temperature and dark matter fluctuations
capable of reproducing all currently available observations.
A seemingly unavoidable outcome of brane inflation, though, is the
production of a network of cosmic strings \cite{jst,costring}, whose 
effects on cosmological observables ranges from negligible to substantial,
depending on the specific brane inflationary scenario \cite{jst2}.

Although cosmic string production towards the end of inflation is 
possible in field theory models \cite{kofman}, the scaling properties 
of the cosmic string networks in brane inflationary scenarios are 
different than that in the familiar 3+1 D
simulations, since intercommutation probabilities are smaller as a
consequence of the existence of extra dimensions. 
In addition to
placing constraints on the amplitude of string induced perturbations of
the CMB, we show that the results place limits on $G\mu/\lambda$,
where $\mu$ is the string tension and $\lambda \le 1$ is related to
the interstring distance as a fraction of the horizon. The value 
of $\lambda$ is sensitive to the intercommutation rate 
of strings, however the exact relation is expected to be model-dependent.

Here we shall first review the essential points of brane inflation,
and examine the constraints imposed by the WMAP observations
if we ignore the contribution from cosmic strings. These constraints
allow us to delineate a range of possible cosmic string tensions.
Then, we assess quantitatively the extent to which cosmic strings
can contribute to the CMB temperature fluctuations and power spectra
of dark matter density perturbations. In this analysis, we hold properties
of the background cosmological model fixed to their best fit values,
as determined by WMAP \cite{wmap_spergel} {\emph without} cosmic strings.
(A more detailed analysis that varies the background cosmology as well
is under way.)
Although the available data favor models without cosmic strings, they
may still allow, within the uncertainties, a contribution from
string-induced perturbations of up to 10\%.
They also imply scalar perturbation indices $n_s(k)$ which, although
still consistent with a broad range of models, may be able to discriminate
among them in future.
We also compute the dark matter density perturbation power spectrum, 
and compare with observational determinations from the 2dFGRS
galaxy survey \cite{2dF}. We discuss the interpretation of these
results in terms of the string tension and efficiency with which the
cosmic string network decays via intercommutation of string segments,
which is reduced in a Universe with extra dimensions. Finally, we
discuss the prospects for detecting B-mode polarization, which is
expected to be a prominent signature of a cosmic string network, in
view of the constraints implied by our analysis.

\section{Brane Inflation and Cosmic String Properties}
\label{braneinf}

Recently, the brane world scenario suggested by superstring theory
was proposed, where the standard model of the strong and electroweak
interactions are open string (brane) modes while the graviton and the
radions are closed string (bulk) modes.
The relative brane positions (i.e., brane separation) in the 
compactified dimensions are scalar fields that have just the right 
properties to act as inflatons.
Thus, the brane inflation scenario emerges naturally in the 
brane world \cite{dvali-tye}. In this picture, 
the inflaton potential is due to
the exchange of closed string modes between branes; 
this is the dual of the one-loop partition function of the open
string spectrum, a property well-studied in string theory \cite{Polchinski1}.
This interaction is of gravitational strength, resulting in a very 
weak (flat) potential, ideally tailored for inflation.

The potential is essentially dictated by the attractive gravitational
(and the Ramond-Ramond) interaction between branes.
As the branes move towards each other, slow-roll exponential inflation
takes place. This yields an almost scale-invariant power spectrum for
density perturbation, except there is a slight red-tilt (at a few
percent level). As they reach a distance around the string scale, the
inflaton potential becomes quite steep so that the slow-roll condition
breaks down. At around the same time, a complex tachyon appears, so 
inflation ends rapidly as the tachyon rolls down its potential. In effect,
inflation ends when the branes collide, heating the universe to
start the standard big bang phase of cosmological expansion. 
This brane inflationary scenario may 
be realized in a number of ways \cite{collection,jst}. The scenario 
is simplest when the radion and the dilaton (bulk) modes are assumed to be
stabilized by some unknown non-perturbative bulk dynamics at the
onset of inflation.
Since the inflaton is a brane mode, and the inflaton potential
is dictated by the brane mode spectrum, it is reasonable to assume
that the inflaton potential is insensitive to the details of the
bulk dynamics.

Coupling of the tachyon to inflaton and standard model fields can allow
efficient heating of the universe  if certain conditions on the
coupling of the tachyon to standard model particles are met \cite{stw}.
As the tachyon rolls down its potential, besides heating the universe,
the vacuum energy also goes to the production of defects, in
particular, cosmic strings. The effect of the resulting cosmic 
string network  may be negligible or rather substantial, 
depending on the particular brane inflationary scenario \cite{jst2}.
However, in all cases, we expect the density perturbation power spectrum in 
CMB to be dominated by the adiabatic fluctuations arising from quantum
fluctuations of the inflaton during brane inflation, not by the nonadiabatic
contributions from cosmic strings.  However,
the contribution to the density perturbation power spectrum in
CMB coming from the cosmic string network may be large enough to
be observable.

We devote this section to a review of the implications of brane inflation.
For a broad set of models, we present results for the slow roll evolution,
fluctuation spectra, string mass scale, and associated cosmic string
tension. (These results follow directly from the treatments in Refs.
\cite{collection,jst, burgess,costring,jst2}.)
We consider the collision of a $\Dp$ brane with a $\Dp$ brane at
an angle $\theta$; collision with a $\Dp$ antibrane corresponds
to $\theta=\pi$. (Here and throughout this section,
we follow \cite{jst}, which contains more details and discussion).
Of the ten spacetime dimensions,
one is the time, three are the large spatial dimensions we live in,
and the rest are compactified. 
Of the compact dimensions, $p-3$ are parallel to
the brane, and we take their compactification lengths to be
$\lpl=2\pi\rpl$, implying a volume $\Vpl=\lpl^{p-3}$. Of the
remaining $d=9-p$ dimensions, we take $d-\dper$ to be compactified
with a size $2\pi/M_s$, where $M_s$ is the string scale, while the
remaining $\dper$ are compactified with a size $\lp=2\pi\rp>
2\pi/M_s$. The 10-dimensional gravitational coupling constant is
\be
\kappa^2=8\pi G_{10}={g_s^2(2\pi)^7\over 2M_s^8}
\label{kappadef}
\ee
where $g_s$ is the expectation value of the dilatonic string coupling,
which is related to the standard model gauge coupling $\alpha(\rpl)$
on a scale $1/\rpl$ by
\be
g_s=2(M_s\rpl)^{p-3}\alpha(\rpl)~;
\label{gsdef}
\ee
the 4-dimensional Planck scale $M_P=(8\pi G)^{-1/2}$ is then
\be
g_s^2M_P^2={M_s^2(M_s\rp)^{\dper}(M_s\rpl)^{p-3}\over\pi}~.
\label{planckdef}
\ee
The outcome of brane inflation will therefore depend on several
parameters, $p$, $\dper$, $\rp$, $\rpl$ and $\alpha(\rpl)$.

We will distinguish between two different potentials for the
interaction between branes, depending on their separations.
(See  \cite{burgess,rabadan,collection}.) For some scenarios,
a fixed lattice of branes is considered to be spread throughout the
compactified dimensions, with a moving brane placed inside one 
lattice square.
At separations small compared to the lattice size of the 
compactification topology, the interaction is ``Coulombic'',
with a potential of the form $V(y)=V_0-U/y^{\dper-2}$ for a separation
$y$ in the large compact dimensions. This potential is suitable for 
inflation resulting from 
the collision of a pair of relatively nearby branes 
at a small angle \cite{rabadan}. When the separation is nearly
equal to the lattice size, an expansion about zero displacement
from the anti-podal point gives $V(y)=V_0-U^\prime y^\sigma$, where
$\sigma$ depends on the compactification topology. This potential 
is suitable for the brane-anti-brane scenario (which corresponds to 
branes at angle $\pi$). In the next two
subsections, we summarize the inflation scenario for interbrane
potentials of these two general forms.

\subsection{Coulombic Inflation}
\label{coulombic}

Consider a potential of the form
\be
V(\psi)=V_0\left(1-{\eta\over(\dper-2)\psi^{\dper-2}}\right)~, 
\ee
with $\psi \propto y$, the interbrane spacing; for the special case $\dper=2$ this
becomes a logarithmic potential,
but the results we derive below may be applied to this special case.
(We only consider $\dper-2\geq 0$ here to simplify our analysis, since the results
generalize easily to the logarithmic case.)
In the slow roll approximation, the equation of motion for $\psi$
becomes
\be
{d\psi\over dL}=-{\eta M_P^2\over\psi^{\dper-1}},
\ee
where $L=\ln a$ is the logarithm of the scale factor $a(t)$, which
we take to be zero at the start of inflation. The
slow roll solution is then
\be
\psi=\left[\psi_i^{\dper}-\dper\eta M_P^2L\right]^{{1\over
\dper}}=\left[\dper\eta M_P^2(\linf-L)\right]^{1\over\dper}
\equiv\left(\dper\eta M_P^2L_r\right)^{1\over\dper}~,
\ee
where the starting value of the field is $\psi_i$, 
the total number of e-folds in inflation, is
\be
\linf={\psi_i^{\dper}\over\dper\eta M_P^2}~,
\label{Ltotalcoul}
\ee
and $L_r=\linf-L$ is the total number of e-folds {\it remaining}
in inflation. The curvature fluctuation spectrum is then
\be
\Deltar={H^4\over 4\pi^2\dot\psi^2}={V_0(\dper
L_r)^{2\left(1-{1\over\dper}\right)}\over 12\pi^2
\eta^{{2\over\dper}}M_P^{2+{4\over\dper}}}
\label{curvcoul}
\ee
where $L_r$ is evaluated when $k/a=H$, or $\ln (k/k_0)=L_{r,0}-L_r$,
where $k_0$ is a reference scale, which crosses with $L_{r,0}$ e-folds
remaining in inflation. The fluctuation spectrum is very flat, with
only slowly varying scalar index, $n_s(k)$:
\ba
n_s(k)-1&=&{d\ln\Deltar\over d\ln k}=-{2\over L_r(k)}
\left(1-{1\over\dper}\right)\simeq
-0.03\left[{60\over L_r(k)}\right]\left(1-{1\over\dper}\right)\nonumber\\
{dn_s(k)\over d\ln k}&=&-{2\over L_r^2(k)}\left(1-{1\over\dper}\right)
\simeq-6\times 10^{-4}\left[{60\over L_r(k)}\right]^2\left(1-{1\over\dper}
\right)~,
\label{nsdnscoulomb}
\ea
both of which are in the range of uncertainty of the determinations in
\cite{wmap_spergel}.

The challenge to this, or any other, inflation model is to have
sufficient inflation as
well as small curvature fluctuation. Since the precise value of
$\linf$ depends on initial conditions as well as parameters of the
model, let us first consider the constraints on the latter implied
by comparing Eq. (\ref{curvcoul}) to the WMAP result
$\Deltarz=2.95\times 10^{-9}A(k_0)$ with $A(k_0)=0.9\pm 0.1$.
To do this, let us consider a particular model with $p=4$ and
a small collision angle, $\theta$;
then we have
\ba
\psi&=&y\sqrt{\tau_4\lpl\over 2}\nonumber\\
V_0&=&{\tau_4\lpl\theta^2\over 4}\nonumber\\
{\tau_4\lpl\over 2}&=&{M_s^4\over 32\pi^3\alpha(\rpl)}\nonumber\\
\eta&=&
{\beta(\dper)\over 16\pi}
~\theta~M_s^{6-\dper}~\left({\tau_4\lpl\over 2}\right)^{\dper-4\over 2}
\nonumber\\
\Deltar&=&{\left(\theta\dper L_r\right)^{2\left(1-{1\over\dper}\right)}
\over 24\left[64\beta(\dper)\right]^{2\over\dper}
\pi^{2+{10\over\dper}}
[\alpha(\rpl)]^{4\over
\dper}}~\left({M_s\over M_P}\right)^{2+{4\over\dper}}~,
\ea
where
\ba
\tau_p & = &\frac {M_s^{p+1}} {(2\pi)^p g_s} \\
\beta &=& \left \{ 
\begin{array}{cc}
{{1 \over { 2 \pi^{\dper/2}}} {\Gamma \left ({d_{\perp} -2} \over 2 
\right)}} & d_{\perp} > 2 \\ {1 \over {\pi}} & d_{\perp} = 2.
\end{array} \right .
\ea
Let us consider the specific example $\dper=2$; for this case we find
\be
\Deltar={\theta L_r\over 768\pi^6[\alpha(\rpl)]^2}\left({M_s\over
M_P}\right)^4\quad\qquad\qquad[\dper=2]~,
\ee
and therefore the string scale is determined to be
\be
{M_s\over M_P}\simeq 2.5\times 10^{-2}[25\alpha(\rpl)]^{1/2}
[A(k_0)]^{1/4}
\left({10\over\theta L_r}\right)^{1/4}
\qquad[\dper=2]
\ee
that is, on the same order of energy as the GUT scale, $10^{15}$ GeV. 
Larger $\dper$ leads to smaller $M_s/M_P$; thus if $\dper=4$ we find
\be
\Deltar={(\theta L_r)^{3/2}\over 12\sqrt{2}\pi^{7/2}\alpha(\rpl)}
\left({M_s\over M_P}\right)^3\quad\qquad\qquad[\dper=4]~,
\ee
which in turn requires
\be
{M_s\over M_P}\simeq 1.6\times 10^{-3}[25\alpha(\rpl)]^{1/3}
[A(k_0)]^{1/3}\left({10\over\theta L_r}\right)^{1/2}\qquad
[\dper=4]~.
\ee
The total number of e-folds in inflation is
\ba
\linf&=&{(M_sy_i)^{\dper}M_s^2\over 64\pi^5\beta(\dper)\theta
[\alpha(\rpl)]^2M_P^2}
={\pi^{\dper(\dper-1)\over\dper+2}(2\zeta_iM_s\rp)^{\dper}\over
\theta^{3\dper\over\dper+2}(\dper L_r)^{2(\dper-1)\over\dper+2}}
\left[{3\Deltar\over 8\beta(\dper)\alpha^2(\rpl)}
\right]^{\dper\over\dper+2}\nonumber\\
&\simeq&{0.025[A(k_0)]^{1/2}\over [25\alpha(\rpl)](10\theta)}\left({10
\over\theta L_r}\right)^{1/2}(\zeta_iM_s\rp)^2\qquad[\dper=2]
\nonumber\\
&\simeq&{0.025[A(k_0)]^{2/3}\over[25\alpha(\rpl)]^{4/3}(10\theta)}
\left({10\over\theta L_r}\right)(\zeta_iM_s\rp)^4
\qquad[\dper=4]~,
\ea
where we have let $y_i=2\pi\rp\zeta_i$ with $\zeta_i\lesssim 1$.
To get $\linf\gtrsim 60$, we must require
 $\zeta_iM_s\rp\gtrsim 50$ for
$\dper=2$ or $\zeta_iM_s\rp\gtrsim 10$ 
for $\dper=4$. Note, though, that for large $\theta$, it is not 
possible to have enough expansion during inflation. In this case,
the images of one brane exert non-trivial forces on the other brane, 
resulting in a power-law type potential.

\subsection{Power Law Inflation}
\label{powerlaw}

Next, we consider potentials of the form
\be
V(\psi)=V_0\left(1-\eta\psi^\sigma\right)~;
\label{plpot}
\ee
such potentials arise for a brane situated near the origin. The
value of $\sigma$ depends on the compactification topology. For
hypercubic compactification, $\sigma=4$, whereas in other cases,
$\sigma=2$. Note that in actuality the potential need not depend
just on interbrane separation in such a picture, and the trajectory
of the brane can be complicated. Here, though, we confine ourselves
to simple one dimensional (diagonal) brane motion.

Following Eq. (\ref{plpot}), we see that the origin -- $\psi=0$ -- is
an unstable equilibrium point, and any perturbation away from it
will result in slow motion of the brane. For $\sigma>2$, the slow
roll solution is
\be
\psi=[\psi_i^{\sigma-2}-\sigma(\sigma-2)\eta M_P^2L]^{1\over\sigma-2}
=[\sigma(\sigma-2)\eta M_P^2L_r]^{1\over\sigma-2}~,
\label{plsoln}
\ee
and the total number of e-folds in inflation is
\be
\linf={\psi_i^{\sigma-2}\over\sigma(\sigma-2)\eta M_P^2}
,
\label{plltot}
\ee
where $\psi_i$ is the starting value for the inflaton. Quantum
fluctuations will imply $\psi_i=\zeta_iH/2\pi$, where $\zeta_1
\sim 1$. The curvature fluctuation spectrum is
\be
\Deltar={V_0(\sigma-2)^2[\sigma(\sigma-2)\eta]^{2\over\sigma-2}
M_P^{2(4-\sigma)\over\sigma-2}L_r^{2(\sigma-1)\over\sigma-2)}\over 12\pi^2}
~.
\label{plfluc}
\ee
The implied fluctuation spectrum is acceptably flat:
\ba
n_s(k)-1={d\ln\Deltar\over d\ln k}&=&-{2(\sigma-1)\over(\sigma-2)L_r(k)}
\simeq -{0.03(\sigma-1)\over\sigma-2}\left[{60\over L_r(k)}\right]\nonumber\\
{dn_s(k)\over d\ln k}&=&-{2(\sigma-1)\over L_r^2(k)(\sigma-2)}
\simeq -6\times 10^{-4}\left({\sigma-1\over\sigma-2}\right)
\left[{60\over L_r(k)}\right]^2~.
\label{nsdnspl}
\ea
For $\sigma=4$, Eqs. (\ref{plfluc}) and (\ref{plltot}) become
\ba
\Deltar&=&{8\eta V_0L_r^{3/2}\over 3\pi^2}\nonumber\\
\linf&=&{\zeta_i^2V_0\over 96\pi^2M_P^4(\eta V_0)}~;
\ea
the observational constraints on the curvature fluctuation spectrum
therefore require
\be
\eta V_0={3\pi^2\Deltar\over 8L_r^{3/2}}\simeq
2.5\times 10^{-11}A(k_0)\left({60\over L_r}\right)^{3/2}~,
\ee
i.e. the potential must be extremely flat. This requirement
is well-known from studies of new inflation, which sometimes
idealize the potential to Eq. (\ref{plpot}) with a small
dimensionless parameter $\lambda$ equivalent to $\eta V_0$.
In Ref. \cite{jst}, a particular toroidal compactification
is proposed where this small parameter is ($F$ is a geometrical
factor related to the compactification geometry)
\be
\eta V_0\simeq {g_s\theta^4F\beta\over 16\pi\alpha^3}
\left({M_s\over M_P}\right)^4~,
\ee
which can be small enough for $\theta\sim 0.1$ provided that
\be
{M_s\over M_P}\simeq 10^{-3}~.
\ee
In this picture, the flatness of the effective potential is
attributed to a relatively small value of the string scale compared
with the Planck mass.

Special treatment is required for $\sigma=2$, which is expected
for any non-hypercubic compactification topology. For this case,
the scale factor grows like a powerlaw in time during slow roll:
\be
\psi=\psi_i\left({a\over a_i}\right)^{2\eta M_P^2}
=\psi_f\left({a\over a_f}\right)^{2\eta M_P^2}~,
\ee
where $\psi_f$ and $a_f$ are the values of the field and scale
factor at the end of slow rolling. Since $d\ln\psi/d\ln a=2\eta
M_P^2$, we require $\eta M_P^2\ll 1$ for slow rolling. It is 
easy to see that for this potential, $\ddot\psi/3H\dot\psi
=2\eta M_P^2/3\ll 1$. Slow roll ends, for this potential, only
when $\dot\psi^2/2V_0=2\eta^2M_P^2\psi^2/3\simeq 1$, or
$\psi_f\simeq (\eta M_P)^{-1}\gg M_P$, or when the polynomial
approximation to the potential fails, which happens when the
brane moves a substantial fraction of a lattice spacings.
The total number of e-folds in inflation is
\be
\linf={\ln(\psi_f/\psi_i)\over 2\eta M_P^2}
={\ln(y_f/y_i)\over 2\eta M_P^2}~.
\label{linfquad}
\ee
The curvature fluctuation spectrum for this case is
\be
\Deltar={(H/2\pi\psi_f)^2e^{4\eta M_P^2L_r}\over
4(\eta M_P^2)^2}=
{(H/2\pi\psi_i)^2(a_i/a)^{4\eta M_P^2}\over 4(\eta M_P^2)^2}
={(H/2\pi\psi_i)^2(a_iH/k)^{4\eta M_P^2}\over 4(\eta M_P^2)^2}
~,
\label{flucquad}
\ee
where evaluating at horizon crossing implies that $k=Ha$, which has
been used to get the final form of the spectrum.
In this case,
\be
n_s-1={d\ln\Deltar\over d\ln k}=-4\eta M_P^2~,
\label{nstwo}
\ee
which is independent of $k$. The WMAP analysis implies that
$\eta M_P^2\lesssim 0.01$. From the first form of Eq. (\ref{flucquad}),
and $\eta M_P^2\simeq 0.01$, it follows that the observed temperature
fluctuations can be accounted for if
($V_0=2\tau_4\ell_\parallel$)
\be
{H\over 2\pi\psi_f}\sim (2\pi M_P y_f)^{-1}\sim 10^{-7}
\left({\eta M_P^2\over 0.01}\right)~,
\ee
in which case $\linf\sim 10^3$. For $y_f\simeq 2\pi\rp$, this
relation implies
\be
{M_s\over M_P}\sim 10^{-6}\left({\eta M_P^2\over 0.01}
\right)~M_s\rp~,
\ee
which is generally smaller than our previous estimates unless $M_s\rp
\sim 10^3$.

\subsection{Cosmic String Properties}
\label{stringprop}

Because the inflaton and the ground state open string modes responsible
for defect formation are different, and the ground state open
string modes become tachyonic and develop vacuum expectation values
only towards the end of the inflationary epoch, various types of defects
(lower-dimensional branes) may be formed.
A priori, defect production after inflation may be a
serious problem. Fortunately, it is argued in Ref.\cite{jst,costring}
that, 
from the properties of superstring/brane theory and the
cosmological evolution of the universe, the only defects copiously
produced are cosmic strings.
In superstring theory, D$p$-branes come with either
odd $p$ (in IIB theory) or even $p$ (in IIA theory). The collision of a
D$p$-brane with another D$p$-brane at an angle (or with an anti-
D$p$-brane) yields D$(p-2)$-solitons (i.e., codimension 2).
Topologically, a variety of defects may be produced. Because they
have even codimensions with respect to the branes that collide, they
have specific properties \cite{Sen1}.
Cosmologically, since the compactified dimensions tangent to
the brane are smaller than the Hubble size, the Kibble mechanism
works only if all the codimensions are tangent to the uncompactified
dimensions. As a consequence, only cosmic strings may be copiously
produced \cite{jst,costring}.

The observational imprint of cosmic strings is determined primarily
by the product of Newton's constant and the cosmic string tension,
$G\mu$, assuming the evolution of the
string network can reach the scaling regime. The value of $\mu$
implied by superstring cosmology depends on several parameters, but
is most sensitive to the string scale, $M_s$.
To get an order of magnitude estimate, we may use the
small $\theta$ case, which is arguably the most likely inflationary
scenario.

The cosmic strings may be D$1$-branes, but most likely, they
are D$(p-2)$-branes wrapping around $(p-3)$-cycles in the compactified
dimensions. If the D$1$-brane is the cosmic string (i.e., $p=3$), its
tension is simply the cosmic string tension:
\be
        \mu = \tau _1 = {M_s^2\over 2 \pi g_s}~.
\ee
However, we expect the string coupling generically to be
$g_s {\ \lower-1.2pt\vbox{\hbox{\rlap{$>$}\lower5pt\vbox{\hbox{$\sim$}}}}\
}1$. (It is well-known that radion and dilaton moduli are not 
stabilized by perturbative dynamics in string theory. Presumably, any superstrongly coupled 
string model is dual to a weakly coupled one, and thus cannot stabilize the moduli either. We therefore expect a moderately strong string coupling, since only in this case will we find non-trivial dynamics.) 
To obtain a theory with a weakly coupled sector in the low energy
effective
field theory (i.e., the standard model of strong and electroweak
interactions with weak gauge coupling constant $\alpha$), it then seems
necessary to have the brane world picture, in which we have
the D$p$-branes for $p>3$, where the $(p-3)$ dimensions are 
compactified to volume $\Vpl$.
Now the cosmic strings are D$(p-2)$-branes, with the $(p-3)$ dimensions
compactified to the same volume $\Vpl$. Noting that a D$p$-brane has
tension
$\tau_p = M_s^{p+1}/(2 \pi)^{p} g_s$,
the tension of such cosmic strings is
\be
\mu = \frac{M_s^{p-1} \Vpl}{(2 \pi)^{p-2} g_s} = \frac{M_s^2 \vpl}{ 2 \pi
g_s} = \frac{M_s^2}{4 \alpha \pi} \simeq 2 M_s^2
\ee
for $\alpha\simeq\alpha_{GUT}\simeq 1/25$.
For one pair of branes at angle $\theta$, only this type of 
cosmic strings is produced topologically.
For a large enough stack of branes colliding, the D$1$-branes 
may also be allowed topologically, but they are
not produced cosmologically. Thus, $\mu\simeq 2M_s^2$ is a
reasonably general estimate. We considered estimates of $M_s$
implied in various scenarios for brane inflation in \S~\ref{coulombic}
and \ref{powerlaw}. These estimates are broadly consistent with
\be
10^{-6}\gtrsim G\mu\gtrsim 10^{-11}~,
\ee
although a smaller range is obtained in any specific model or class
of setups. For example, for branes colliding at a small angle,
a likely range is
\be
5\times 10^{-7}\gtrsim G\mu\gtrsim 7\times 10^{-8}~.
\ee
Thus, brane inflation can lead to cosmic string tensions below, but
not far below, current observational bounds.

\subsection{Tensor modes}
\label{tensor}

During slow roll, the tensor power is
\be
\Delta_h^2(k)={128G^2V_0\over 3}={2V_0\over 3\pi^2 M_P^4}
\ee
and is smaller than the scalar power by the factor
\be
r(k)=8M_P^2\left({V^\prime\over V}\right)^2={8\over M_P^2}
\left({d\phi\over d\ln a}\right)^2~.
\ee
How small $r(k)$ is depends on the specific brane inflation
model. For branes intersecting at an angle $\theta$ we find
that
\ba
\Delta_h^2(k)&=&{2V_0\over 3\pi^2M_P^4}={\theta^2M_s^4\over 96\pi^5
\alpha(\rpl)M_P^4}\nonumber\\
&=&3.3\times 10^{-12}~(10\theta)^2~[25\alpha(\rpl)]
\left[{10A(k_0)\over\theta L_r}\right]\qquad\qquad\quad[\dper=2]
\nonumber\\
&=&5.4\times 10^{-17}(10\theta)^2[25\alpha(\rpl)]^{1/3}[A(k_0)]^{4/3}
\left({10\over \theta L_r}\right)^2\qquad\quad[\dper=4]~.
\ea
In this case, the amplitude of the scalar mode is smaller than
the amplitude of the perturbations due to cosmic strings by a small
numerical factor times $\theta^2$, unless cosmic string intercommutation
is extremely inefficient; see \S~\ref{stringprop} and \ref{calcinterp}
below. For powerlaw brane-antibrane inflation, $\theta\to\pi$, and
$V_0=M_s^4/(2\pi)^3\alpha$ \cite{burgess,jst}, so for this case
we find
\be
\Delta_h^2(k)={M_s^4\over 12\pi^5\alpha M_P^4}~.
\ee
Nominally, these perturbations can be comparable to those induced by
cosmic strings, although they may be relatively suppressed by the small
numerical factor $(12\pi^5\alpha)^{-1}\simeq 0.007/(25\alpha)$. However,
the spectrum of fluctuations produced by cosmic strings will still 
distinguish them from those due to primordial tensor modes.
Both strings and the primordial tensor modes result in the B-type 
polarization
of the CMBR. The predicted angular power spectrum, $C_l^{BB}$, has been
calculated for tensor modes from inflation (see e~.g. \cite{alessandro}).
It has a generic feature that most of the power is on larger angular 
scales, in the region $l\lesssim 100$.
This is very different from the shape of the $C_l^{BB}$ spectrum 
predicted by cosmic strings. There the dominant contribution comes 
from the vector modes and,
as one can see from Fig.~\ref{fig:clbb}, most of the power is on smaller
scales: $700\lesssim l \lesssim 1000$.

As of today, the B-type polarization has not been detected
\cite{kovac} and the 
experimental constraint on $r(k)$ is rather mild:
$r(k_0=0.002{\rm Mpc}^{-1})\lesssim 0.71$ \cite{wmap_peiris}.


\section{CMB, Matter Density and Polarization Power Spectra:
Calculation and Interpretation}
\label{calcinterp}

The fluctuations expected to arise from brane inflation should be an
incoherent superposition of contributions from adiabatic perturbations
initiated by curvature fluctuations $\Deltar$ and active perturbations
induced by the decaying cosmic string network. For example, the resulting
CMB temperature maps will yield 
\be
C_l=WC_l^{\rm adiabatic}+BC_l^{\rm strings},
\label{clsuperpos}
\ee
were $W$ and $B$ are weighting factors. Analogous expressions hold for
matter density and polarization power spectra. In Eq. (\ref{clsuperpos}),
the weight factors $W$ and $B$ determine the relative importance of the
adiabatic and cosmic string contributions. We choose the weight factor
$W$ so that $W=1$ when there are no cosmic strings.

In computing the combined effects of adiabatic and cosmic string
perturbations, we have kept the cosmological background parameters
fixed at their best-fit values according to \cite{wmap_spergel}. In addition
to $B$ and $W$, we vary the spectral index of the scalar
curvature fluctuations, $n_s$. The tensor contribution to the adiabatic
component was set to zero, since, as discussed in \S~\ref{tensor},
it is likely to be small. When fitting to both WMAP and the 2dFGRS data, we
considered two cases (described in more detail in \S~\ref{methods}): with bias $b$ fixed,
and with $b$ being an additional parameter of the fit.

\subsection{Cosmic strings and the CMBR}
\label{model}

Perturbations due to cosmic strings were calculated
using the model first introduced in \cite{aletal} and further
developed in \cite{levon,gangui}. The main idea is to represent the cosmic
string network by a collection of uncorrelated, straight
string segments moving with random, uncorrelated velocities. All segments are
produced at some early epoch and, at every subsequent epoch, a certain
fraction of the number of segments decays in a way that maintains
network scaling.  The length of each segment at any time is taken to
be equal to the correlation length of the network which, together with
the root mean square velocity of segments, are computed from the
velocity-dependent one-scale model of Martins and Shellard
\cite{MS96}. The positions of segments are drawn from a uniform
distribution in space and their orientations are chosen from a uniform
distribution on a two-sphere.

This model is a rather crude approximation of a realistic
string network. However, with a suitable choice of model parameters,
its main predictions for CMB and matter power spectra have been shown to
be in agreement with results obtained using other local string sources
\cite{joao,copmagst}. The main advantage of our model is its flexibility.
For example, parameters can be chosen to describe stings with
different scaling properties, different amounts of small scale structure etc.
This is especially valuable when describing strings produced in
brane inflation, since strings are expected to intercommute with a
lower probability in the presence of extra spatial dimensions \cite{jst2}.

It is well known that properties and possible observational signatures
of global and local strings can be dramatically different. Global strings
predict almost no power on small angular scales for CMB temperature
anisotropy \cite{pen_ps}, while local strings produce a quite significant
broad peak at $l\sim 450$ in a spatially flat universe
\cite{aletal,levon,joao,copmagst,closed}.
Also, local strings develop a signification amount of small-scale
structure over the course of their evolution, which tends to
suppress the vector component of metric perturbations. 
For that reason one would expect the strength of the 
$B$-type polarization \cite{pen_pol} sourced by global strings to
be somewhat larger than that of local strings \cite{newwork}. 




Numerical simulations \cite{BenBouch,AllenShellard90,AlTur89} show
that during the radiation and matter dominated eras the string
network evolves according to a scaling solution, which on
sufficiently large scales can be described by two length scales.
The first scale, $\xi(t)$, is the coherence length of strings,
i.~e. the distance beyond which directions along the string are
uncorrelated. The second scale, $L(t)$, is the average
inter-string separation. Scaling implies that both length scales
grow in proportion to the horizon. Simulations indicate that $\xi
(t) \sim t$, while $L(t) = \gamma t$, with $\gamma \approx 0.8$ in
the matter era \cite{BenBouch,AllenShellard90}. The
one-scale model \cite{Vilenkin81b,Kibble85}, in which the two length
scales are assumed to be equal, has been quite successful in
describing the general properties of cosmic string networks
inferred from numerical simulations. These simulations have
assumed that cosmic strings would reconnect on every intersection.
It is of interest to us, however, to also consider the case when
the reconnection probability is less than one. If strings can move
and interact in extra dimensions then, while appearing to
intersect in our three dimensions, they may actually miss each
other. Hence, the effective intercommutation rate of these strings
will generally be smaller than unity.  As a consequence, one would
expect more strings per horizon in these theories.
Because of the straightening of wiggles on sub-horizon scales due
to the expansion of the universe, one would still expect $\xi(t)
\sim t$. However, the string density would increase, therefore
reducing the inter-string distance. Hence, smaller
inter-commutation probabilities imply smaller $\gamma$.

Consider a string network in a volume $V$ described by the two length
scales introduced above. On average there is one string segment of length $\xi$ 
per volume $\xi L^2$. The rough number of such string segments is
\be
N={V \over \xi L^2} \ .
\ee
If the energy per unit length is $\mu$, then the total energy of
the string network is
\be
E = N \mu \xi = {V \mu \over L^2} \ .
\ee
and the energy density is just $\rho=E/V= \mu/L^2$.

Now suppose we want to calculate the effect of this string
network on the CMB temperature anisotropy. In particular, we want to
find the power spectrum, i.~e. the 2-point function. For simplicity,
let us assume that only $\rho(t)$ affects CMB (in general we would
have to consider all components of the string network's energy-momentum
tensor $T_{\mu}^{\nu}$). Then to evaluate the CMB power spectrum it suffices
to know the 2-point unequal time correlators $\langle\rho(k,t_1)\rho(k,t_2)
\rangle$ at
all $k$, $t_1$ and $t_2$. Here $\rho({\bf k},t)$ is the Fourier transform of
$\rho({\bf x},t)$. The CMB power spectrum is roughly given by
\be
C_l = \int dk \int dt_1 \int dt_2 \ \hat{L}_l(t_1,t_2,k)
\big[ \langle\rho(k,t_1)\rho(k,t_2)\rangle \big] \ ,
\ee
where $\hat{L}_l(t_1,t_2,k)$ is a linear operator.

Again, for simplicity, let us assume that $\xi(t) \sim t$ and $L=\gamma t$ at
all times, namely that the network scales perfectly with time.
We want to see how $C_l$, or equivalently $\langle\rho(k,t_1)\rho(k,t_2)
\rangle$,
depends on $\gamma$ and $\mu$. Let us also assume that the segments are
straight.

At time $t$ there are roughly $N(t)=V/\xi L^2$ segments and
\be
\rho(k,t)= \sum_{i=1}^{N(t)} \rho^{(i)}(k,t)
\equiv \sum_{i=1}^{N(t)} \tilde{\rho}^{(i)}(k,t) \mu \xi(t)  \ ,
\ee
where $\tilde{\rho}^{(i)}(k,t)=\rho^{(i)}(k,t)/(\mu \xi(t))$ to
factor out dependences on $\mu$ and $\xi(t)$ (the phase of
$\tilde{\rho}$ may still depend on $\xi(t)$ but the amplitude does not).
Now we can write:
\be
\langle\rho(k,t_1)\rho(k,t_2)\rangle = \sum_{i=1}^{N(t_1)} \sum_{j=1}^{N(t_2)}
\langle\tilde{\rho}^{(i)}(k,t_1) \rho^{(j)}(k,t_2)\rangle \mu^2 \xi(t_1) \xi(t_2) .
\ee
Because individual segments are statistically independent
\be
\langle\tilde{\rho}^{(i)}(k,t_1) \rho^{(j)}(k,t_2)\rangle =
\delta_{ij} \ \langle\tilde{\rho}^{(i)}(k,t_1) \tilde{\rho}^{(i)}(k,t_2)
\rangle  \ ,
\ee
and therefore
\be
\langle\rho(k,t_1)\rho(k,t_2)\rangle = \sum_{i=1}^{{\rm min}[N(t_1),N(t_2)]}
\langle\tilde{\rho}^{(i)}(k,t_1) \tilde{\rho}^{(i)}(k,t_2)\rangle
 \mu^2 \xi(t_1) \xi(t_2)
\ee
To interpret ${\rm min}[N(t_1),N(t_2)]$ it might help to think that all
segments were there at the initial time but over course of their
evolution some of them decayed.
For certainty, let us assume $t_1<t_2$ and therefore $N(t_1)>N(t_2)$.
We can now write
\be
\langle\rho(k,t_1)\rho(k,t_2)\rangle = \sum_{i=1}^{N(t_2)}
\langle\tilde{\rho}^{(i)}(k,t_1) \tilde{\rho}^{(i)}(k,t_2)\rangle
 \mu^2 \xi(t_1) \xi(t_2) =
N(t_2) \langle \tilde{\rho}^{(1)}(k,t_1) \tilde{\rho}^{(1)}(k,t_2)\rangle \mu^2 \xi(t_1) \xi(t_2) \ ,
\ee
where the last step is possible because all segments are statistically
identical and hence
\be
\langle\tilde{\rho}^{(i)}(k,t_1) \tilde{\rho}^{(i)}(k,t_2)\rangle = 
\langle\tilde{\rho}^{(1)}(k,t_1)
\tilde{\rho}^{(1)}(k,t_2)\rangle \ \ \ {\rm for \ all} \ i.
\ee
Substituting $\xi \sim t$, $L=\gamma t$ and $N(t) = V/(\xi L^2) \sim V/(\gamma^2 t^3)$ 
we find
\be
\langle\rho(k,t_1)\rho(k,t_2)\rangle = {V \over \gamma^2 t_2^3} \mu^2
t_1 t_2 \langle\tilde{\rho}^{(1)}(k,t_1) \tilde{\rho}^{(1)}(k,t_2)\rangle
= {\mu^2 \over \gamma^2} \times F(k,t_1,t_2) \ ,
\ee
where $F(k,t_1,t_2)$ is independent of $\xi$ or $\mu$, and therefore
\be
C_l^{\rm strings} \propto {\mu^2 \over \gamma^2} \ .  
\label{xiscaling}
\ee
Rather than working with the parameter $\gamma=L(t)/t$, we
define a related parameter $\lambda \equiv l(\tau)/\tau$, where $l(\tau)$ is
the comoving interstring distance, $L/a$, and $\tau$ is the conformal
time, $d\tau=dt/a$. In our code we take parameter $\lambda$ to be a constant.
Correspondingly, the meaning of the parameter $B$ is 
\be
\label{bdep}
B = {C_l(\lambda,G\mu) \over C_l(\lambda_0,G\mu_0)} =
\left[{\mu \lambda_0 \over \mu_0 \lambda} \right]^2
\ee
with $G\mu_0 = 1.1 \times 10^{-6}$ and $\lambda_0 = 0.25$ adopted
as reference values.

The relation between parameter $\lambda$ and the string 
intercommutation probability $p$ has been a subject of extensive
research. Qualitative arguments in \cite{DV04} suggest that
$\lambda \propto p^{1/2}$. However, these arguments do not take
into account possible effects of small-scale structure on strings.
Results of the most recent numerical simulations \cite{AS05} suggest
that for $p>0.1$ the scaling string density is effectively the same
as for $p=1$, while for smaller $p$ one should expect $\lambda \propto p^{0.3}$.
Such weak dependence of $\lambda$ on $p$ can be explained by 
multiple opportunities for string reconnection during one crossing time,
due to small-scale wiggles.

\subsection{Methods}
\label{methods}

We have performed a partial
statistical analysis in which we held the parameters of the background cosmological model
(total, matter, baryon and dark energy density parameters, Hubble constant,
reionization optical depth) fixed at their WMAP best fit values according
to \cite{wmap_spergel}. More specifically, we considered a flat $\Lambda$CDM universe with
$\Omega_{\rm CDM}=0.225$, $\Omega_b=0.045$, $\Omega_{\Lambda}=0.73$, $H_0=71$~km/s/Mpc
and $\tau=0.17$. The scalar spectral index $n_s$ was allowed to vary within
bounds set by the prior $0.8\le n_s \le 1.2$, in increments of
$\Delta n_s = 1.25 \times 10^{-3}$.

The string spectra were calculated
only once, using the string model parameters chosen to produce
spectra that roughly agree with \cite{aletal,levon,joao,copmagst}. In particular,
we set $G\mu_0 = 1.1 \times 10^{-6}$, which, if strings were the only source of
inhomogeneity, would result in temperature anisotropy in a rough agreement
with observations on COBE scales. The string intercommutation probability $\lambda$
was approximately $\lambda \sim 0.25$ and was allowed to vary only insignificantly
during the radiation-matter domination transition.

The CMBR and linear matter power spectra for
both adiabatic and string parts were computed using respectively
modified versions of CMBFAST \cite{cmbfast}. However, one cannot directly compare
linear matter spectra outputted by CMBFAST to the galaxy clustering
data published by the 2dFGRS team. We have ``processed'' the theoretical
power spectrum $P^{th}(k)$ for both adiabatic and string components following
a procedure similar to that prescribed in Section 5.1.4 of \cite{wmap_verde}.
First of all, we output $P^{th}(k)$ at the effective redshift of the
2dFGRS survey: $z_{\rm eff}=0.17$ (the valued suggested in \cite{wmap_verde}).
Then we correct for the redshift space distortions using the approximate formula given
in \cite{wmap_verde}:
\be
P^{th}(k) \rightarrow
P^{th}_z(k) = (1 + {2 \over 3} \beta + {1\over 5} \beta^2) P^{th}(k) \ ,
\ee
with $\beta=0.45$. We then convolve it with the 2dF window function \cite{2dF}
using the matrix $M_W$ provided on the 2dFGRS website:
\be
P^{th}_z(k)\rightarrow  P_W(k)= \sum_q M_W(k,q) P^{th}_z(q)
\ee
and, finally, multiply by the bias factor to obtain the spectrum that can be compared
to data:
\be
P(k)=b^2 P_W(k)
\ee
where $b=\Omega_m^{0.6}/\beta$.

We have chosen to fit to the binned version of WMAP data, for both temperature (TT)
and cross temperature-polarization (TE) angular spectra. When fitting to
WMAP+2dF we had a choice of making the bias $b$ an additional parameter
or using a prescribed fixed value. We have investigated both possibilities and
will refer to them the ``$b$-free'' and ``$b$-fixed'' model.

\subsection{Results}
\label{results}

We are interested in constraining the fractional contribution to the
WMAP and 2dFGRS data given by Eq.~(\ref{clsuperpos}). To do this properly,
from first principles, would require using theoretical models with both
cosmic strings and adiabatic perturbations to generate the relevant 
sky maps, and then compare these directly to the data to compute likelihood
functions that can be used, along with appropriate priors, the posterior
probability distributions for the parameters of the models. To get a 
quick and dirty bound on the most interesting parameter, $B$, we have
chosen to adopt a simpler approach in which we treat the published
results for $C_l$ of the TT and TE sky maps, and the 2dFGRS power
spectra as data, and use them and their uncertainties to construct
a $\chi^2$ statistic in the usual way. For each value of $B$, we minimize
the statistic with respect to variations in the parameters $n_s$, $W$,
and the bias parameter, to find $\chi^2_{min}(B)$. 

\begin{figure}[!]
\scalebox{0.420}{\includegraphics{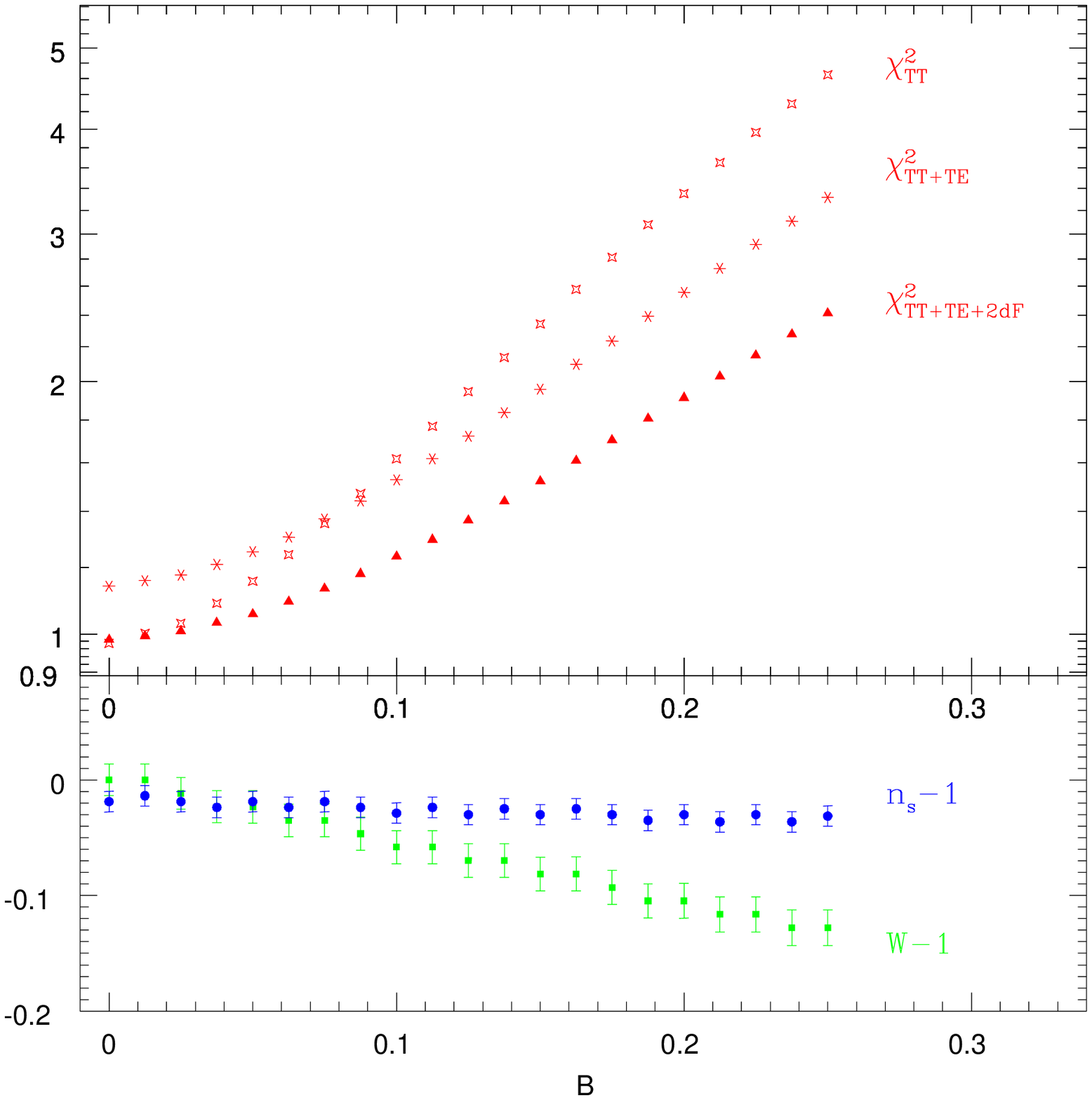}}
\scalebox{0.420}{\includegraphics{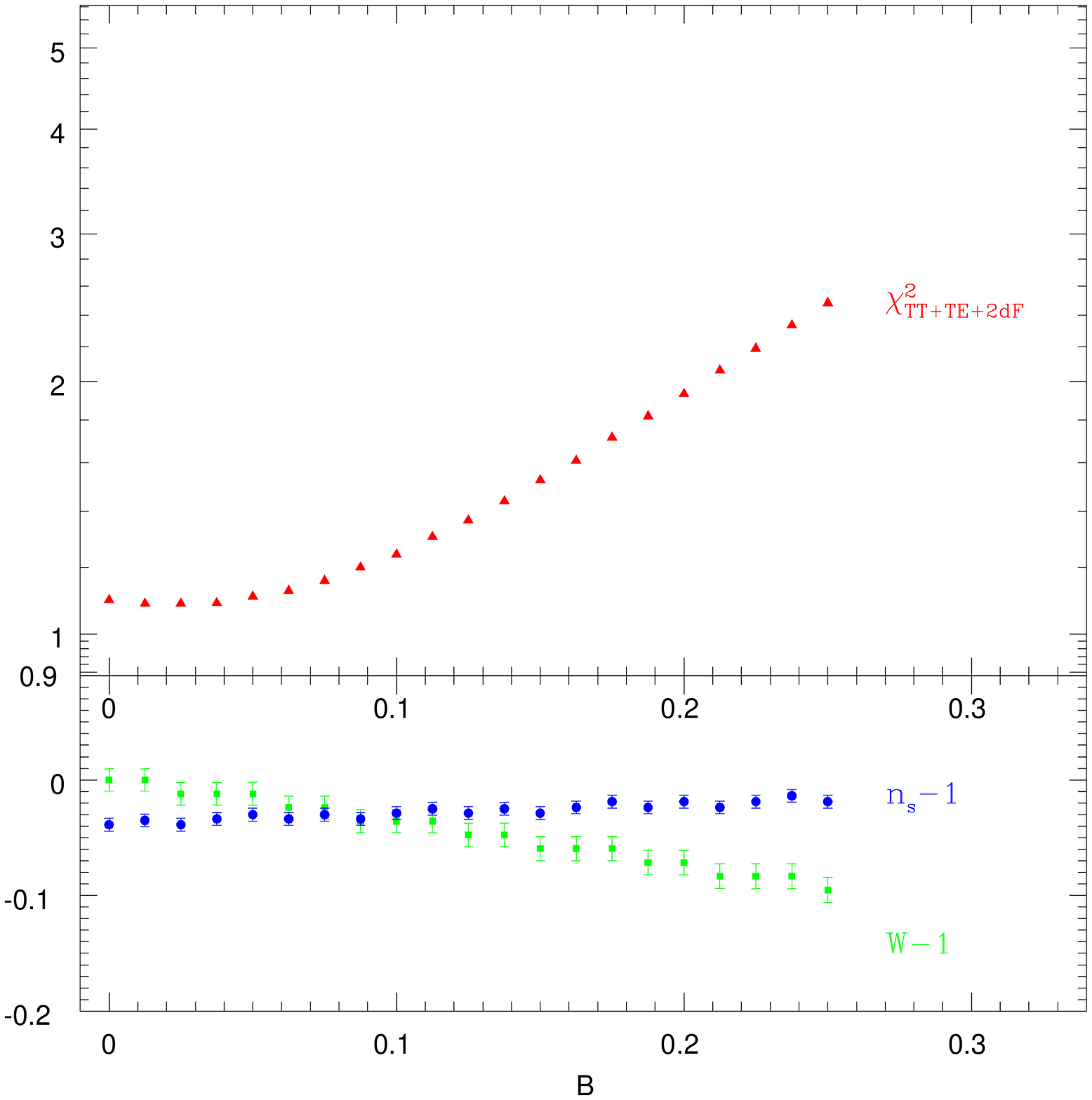}}
\caption{\label{fig:chiminall} (Left) Minimum reduced
$\chi^2/\nu$ as a function of $B$ for the fit
of the $b$-free model to the WMAP's TT (red triangles), TT+TE (red stars)
and TT+TE+2dF (red hollow squares) with corresponding best fit values of
$W-1$ (green squares) and $n_s-1$ (blue circles). We only show $W$ and $n_s$
from the fit to the TT data alone. The error bars correspond to the diagonal
elements of the covariance matrix. Adding the remaining data sets changes the
best fit values of $W$ and $n_s$ by only a small amount, well within the plotted error
bar. (Right) Same as on the left but for the
$b$-fixed model with all data sets included. In the left panel, $\nu=93$ and
in the right $\nu=94$. The values of the reduced $\chi^2/\nu$ at $B=0$ are,
in the left panel, 0.97 (TT), 1.1 (TT$+$TE) and 0.98 (TT$+$TE$+$2dF),
and 1.1 in the right panel.}
\end{figure}

If our ``data'' really represented independent data points, then we could
be confident that at each $B$ the value of $\chi^2_{min}(B)$ would obey
a distribution that roughly has a mean value equal to the number of
degrees of freedom in the fit, $\nu$, and a standard deviation 
$\sqrt{2\nu}$. Fig. \ref{fig:chiminall} shows the results for 
$b$-free (left) and $b$-fixed (right) (as defined at the end
of \S~\ref{methods}). For our procedure, we have $N=38({\rm TT}+
26({\rm TE})+32({\rm 2dF})=96$ ``data'' points. We keep cosmological
paramters such as $\Omega_m$, $\Omega_b$, $\Omega_\Lambda$, $h$
and $\tau({\rm reionization})$ fixed at their best fit values from
\cite{wmap_spergel}. Therefore, we have either three ($n_s$, $W$, and
$b$ for $b$-free evaluation) or two ($n_s$ and $W$ for $b$-fixed
evaluation) parameters in each of the panels in Fig. \ref{fig:chiminall},
so $\nu=93$ ($b$-freee) or $94$ ($b$-fixed), and $\sqrt{2\nu}\simeq
14$ in either case. Since the two panels in the figure actually show
the reduced statistic $\chi^2(B)/\nu$, we could regard values of
$B$ with $\chi^2(B)/\nu$ within approximately $0.15n$ of the minimum
value as representing more or less equally good fits to the data to
within $n$ ``sigma''. Below, we shall use $\chi^2/\nu\lesssim
1.25$ for placing a limit on $B$; this can be interpreted as either
a 1.7``$\sigma$'' bound, or as a crude attempt to account for the
fact that we are also ignoring the effects of varying several more
cosmological parameters. Although assessing the acceptability
of models with different values of $B$ is not justified rigorously,
we note that the minimum values of $\chi^2/\nu$ shown
in either panel of Fig.~\ref{fig:chiminall} are close to one,
which may be taken as {\it a posteriori} assurance that our method
for comparing models with different values of $B$ may not be far off.

The left panel of Fig.~\ref{fig:chiminall}, corresponding to the $b$-free model,
contains plots of the minimized reduced $\chi^2_{\rm TT}/\nu$, $\chi^2_{\rm TT+TE}/\nu$
and $\chi^2_{\rm TT+TE+2dFa/}\nu$ ($\nu=93$) as functions of $B$
computed using, respectively, WMAP's TT spectrum only, TT and TE,
and TT, TE and the 2dF spectrum.
As one can see from that plot, in all cases the lowest value of the minimum
$\chi^2(B)/\nu$ occurs at
$B=0$, i.~e. when there is no contribution from strings. However, as
additional data sets are added to the fit, the values of the minimum
$\chi^2(B)$ become smaller.
Fig.~\ref{fig:chiminall} also shows the values of $n_s-1$ and $W-1$ values
that minimized $\chi^2/\nu$ at each $B$, with estimated uncertainties, of
order $\pm 0.01$ for both.
Not plotted are the values of the bias parameter $b$, which varied from
0.97 at $B=0$ to 1.02 at $B=0.25$.
Note that the value of $n_s-1$ is typically around $-0.02\pm 0.01$ for $B\lesssim
0.1$, the region where the minimum $\chi^2(B)/\nu$ is less than about $1.25$.

The right panel of Fig.~\ref{fig:chiminall} corresponds to the $b$-fixed model,
i.~e. the model in which the value of the bias was held fixed at $b=\Omega_m^{0.6}/\beta$.
Results are shown for the minimum value of $\chi^2(B)/\nu$ constructed using
all three data sets. Here we find that the lowest value of the minimum 
$\chi^2(B)/\nu$ is at $B=0.025$, not $B=0$.
In view of the limitations of our statistical analysis,
one cannot attach much significance to
either the existence of this minimum, or the precise nonzero $B$ at which
it occurs, but it is worth
observing that adding a small string contribution may be welcome when matching
the CMBR normalized matter spectra to the galaxy data.
Moreover, this result shows that at the very least we cannot exclude values
of $B\sim 0.025$.
We note that the values of $n_s-1$ corresponding to the minimum $\chi^2$ per
degree of freedom grow with $B$ in this panel and,
for $B\lesssim 0.1$,  they fall within the range
$n_s-1 = -0.03\pm 0.01$.

\begin{figure}[!]
\scalebox{0.420}{\includegraphics{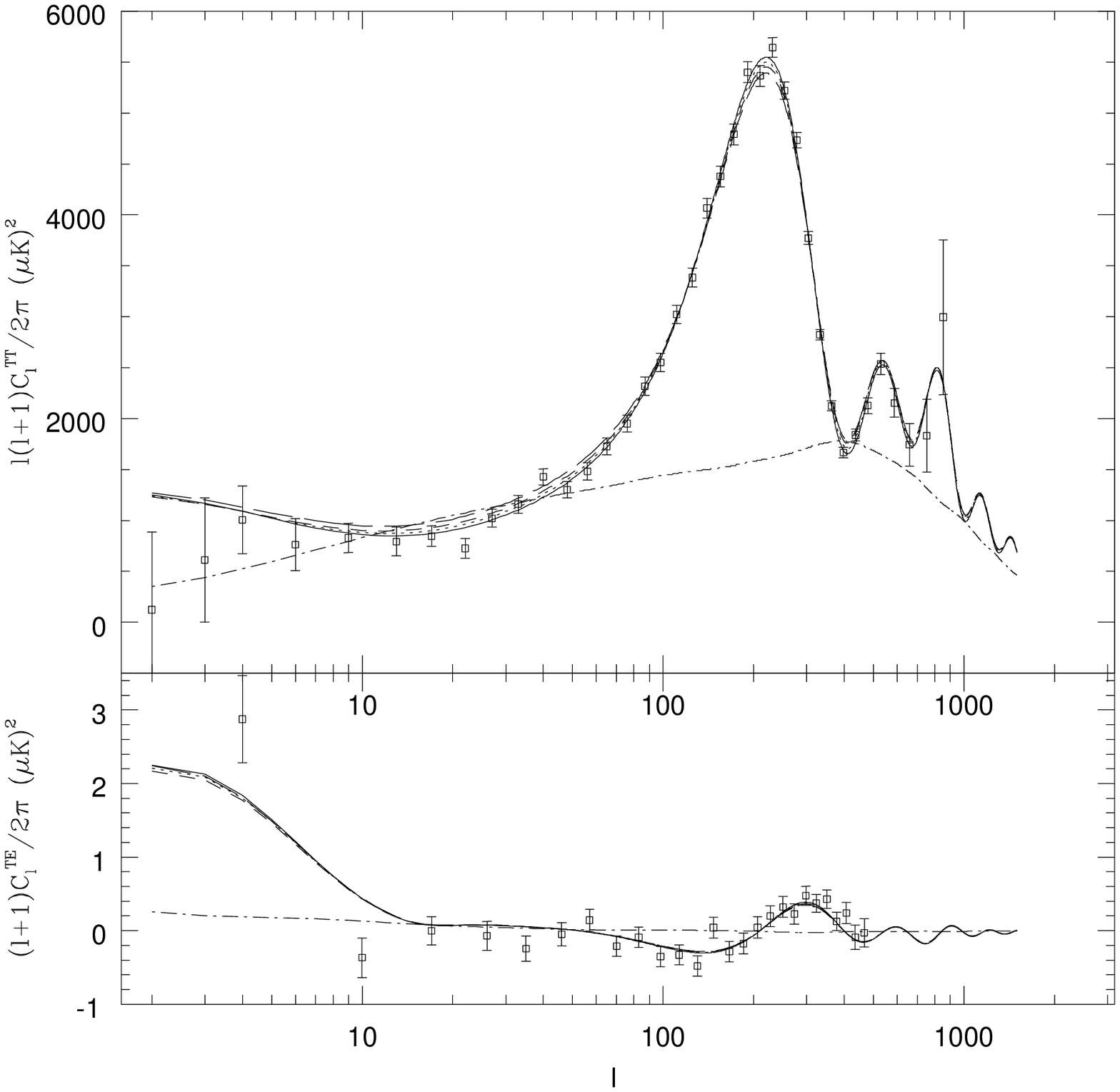}}
\scalebox{0.420}{\includegraphics{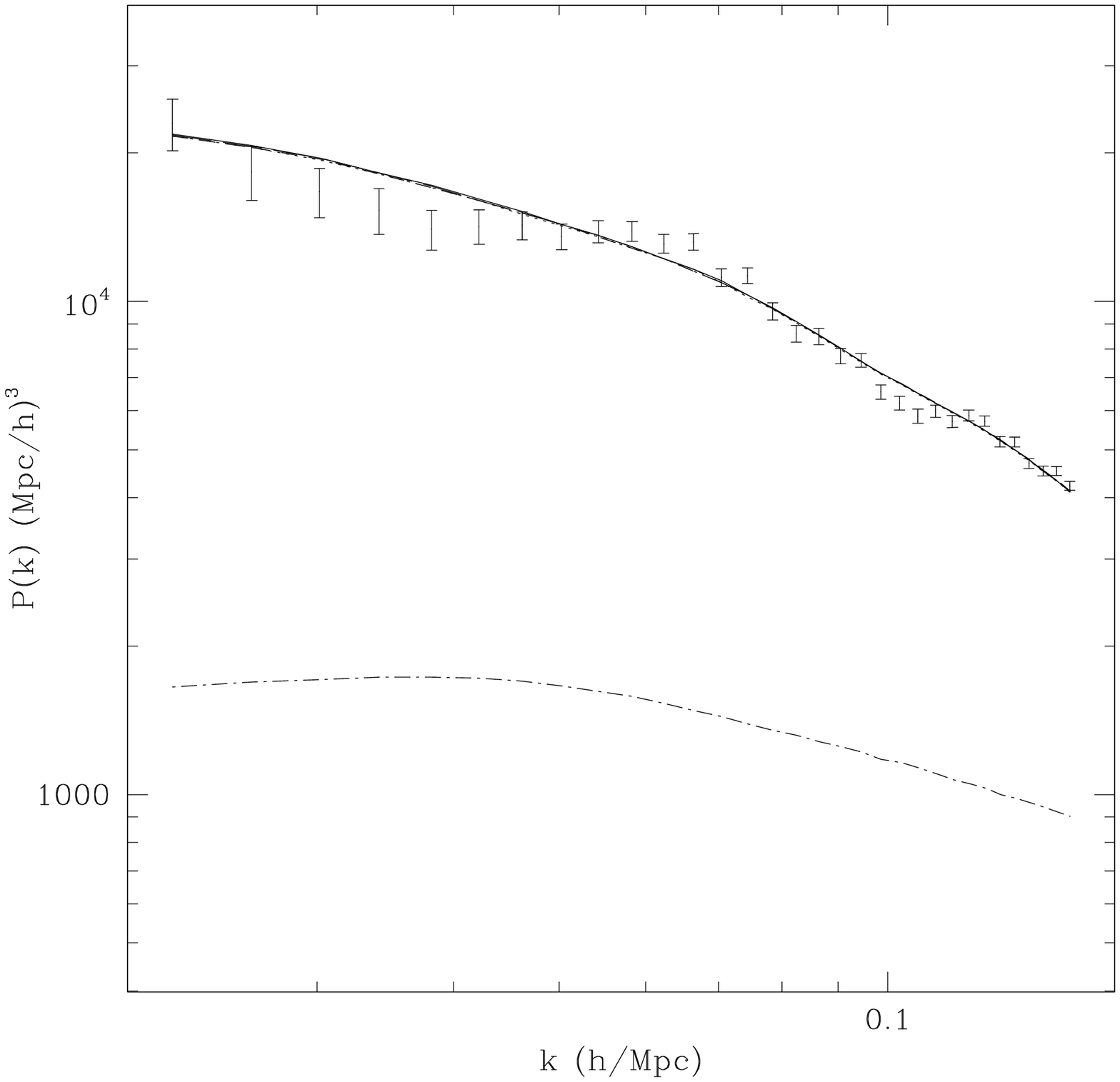}}
\caption{\label{fig:free}
(Left) Plots of the best fit $C_l^{TT}$ and $C_l^{TE}$ computed
using the $b$-free model for different values of $B$ together with the WMAP's binned data.
(Right) Corresponding plots of the galaxy clustering power spectra together with
the 2dFGRS data. On both plots, the solid line corresponds to $B=0$, dotted line - $B=0.05$,
short dash line - $B=0.1$, long dash line - $B=0.15$ and the dash-dot line corresponds to
the pure string contribution, i.e. $(W=0,B=1)$.}
\end{figure}

\begin{figure}[!]
\scalebox{0.420}{\includegraphics{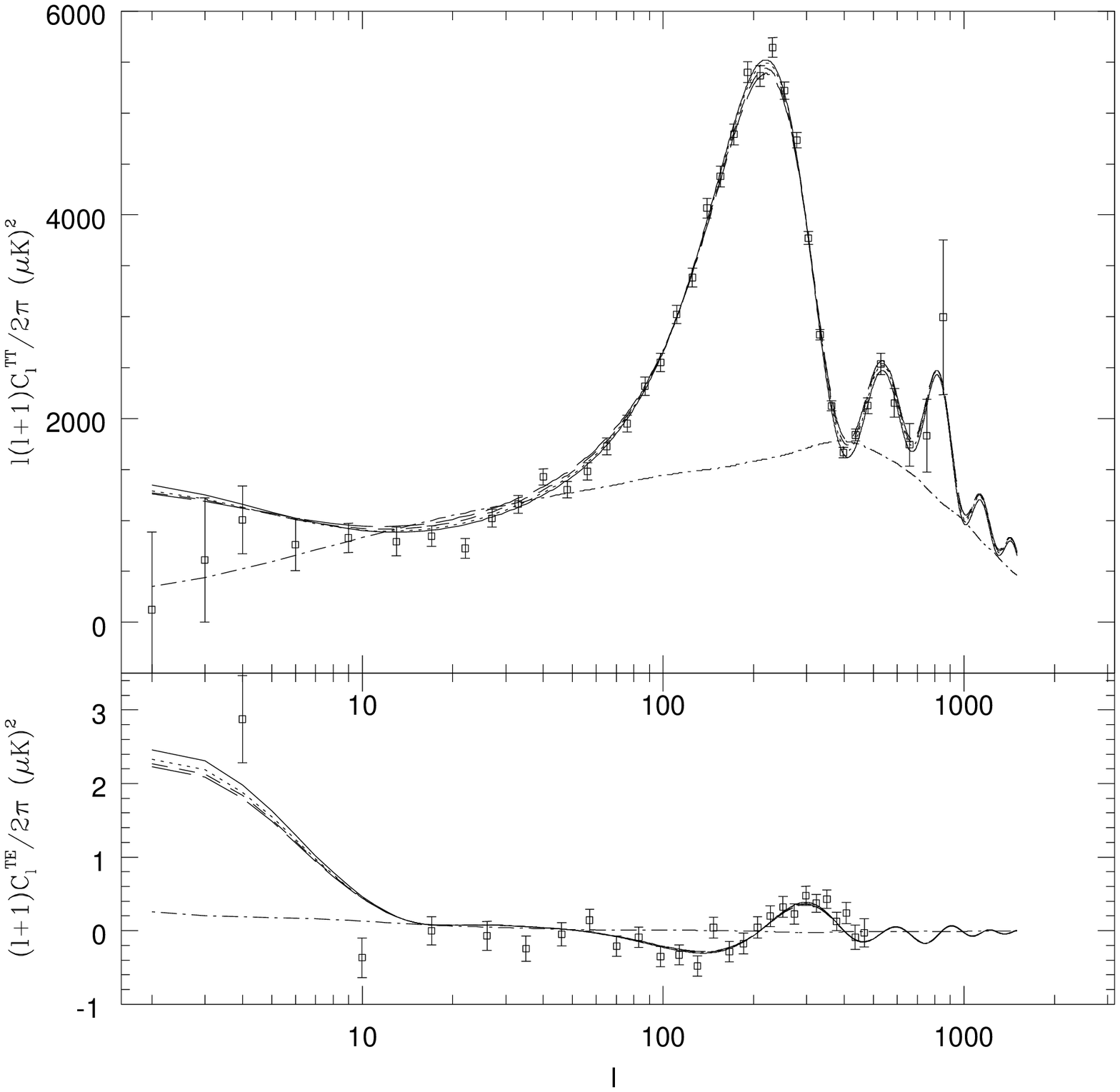}}
\scalebox{0.420}{\includegraphics{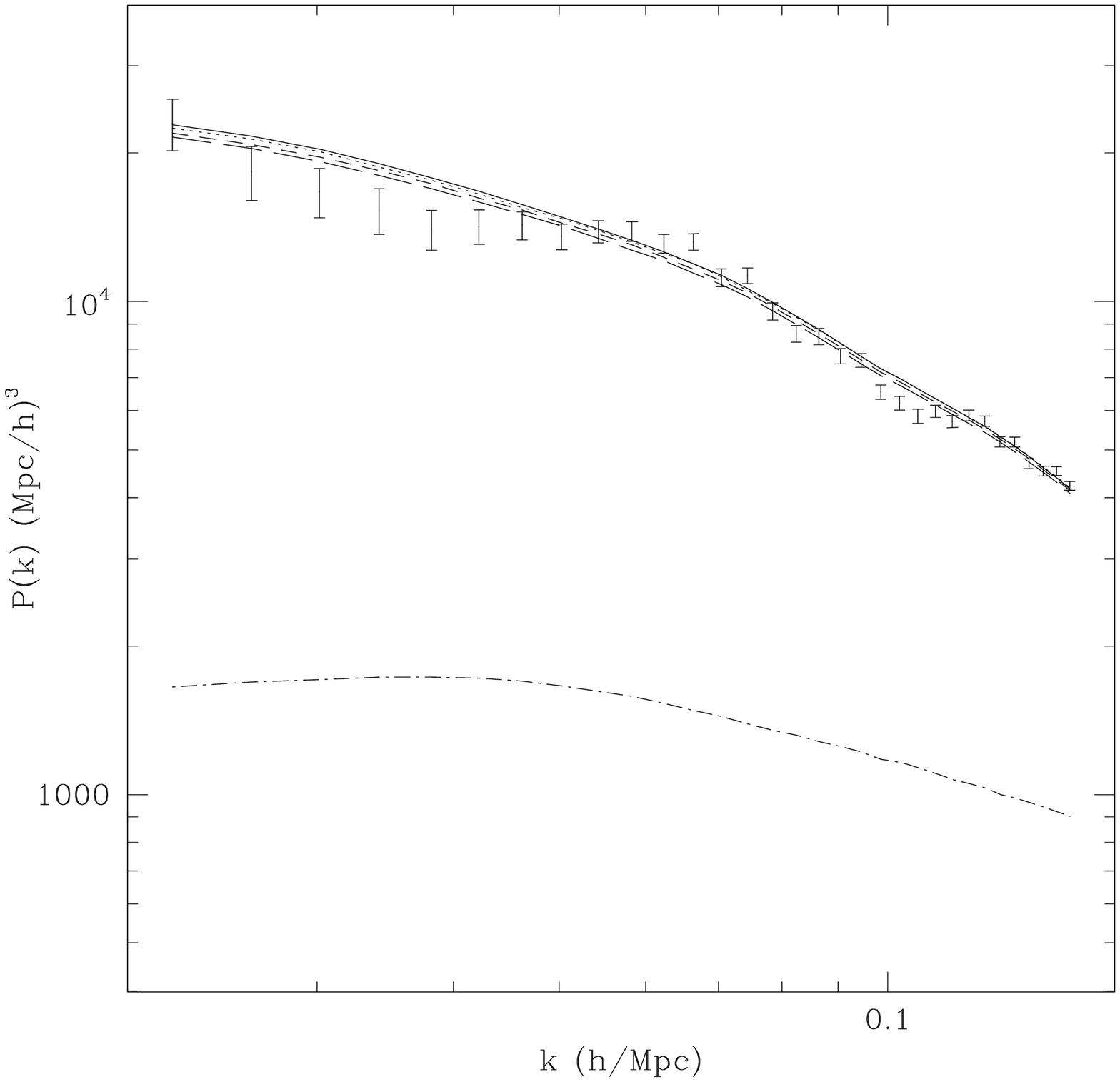}}
\caption{\label{fig:fixed} Same as in Fig.~\ref{fig:free} but for the $b$-fixed model.}
\end{figure}

In Figures \ref{fig:free} and \ref{fig:fixed} we plot the adiabatic
plus string CMB and matter power spectra superposed using Eq.~(\ref{clsuperpos})
with different values of $B$ and corresponding best fit values of $n_s$ and $W$
for the $b$-free and $b$-fixed models. In both figures, the solid line corresponds
to $B=0$ and the dash-dot line to $(W=0,B=1)$.


The simple adiabatic $\Lambda$CDM model with a constant $n_s$ fits the WMAP data
very well in general, but much worse so on very large scales.
For the first few multipoles, $2\leq l\leq 6$, the experiment finds a clear deficit of
power, as compared to the prediction of the model. One could hope
that, because cosmic strings generically predict such lack of power for
low multipoles, adding a string component would improve the fit on large scales.
However, as one can see from Figs.~\ref{fig:free} and \ref{fig:fixed},
this is not the case. The reason is that cosmic strings fit the data
extremely poorly in the region of the first and the second acoustic peaks, where
the WMAP error bars on $C_l$s are {\it very} small. The ``benefit'' from adding
strings on larger scales, where the error bars are large, is offset by the much larger
``damage'' that strings cause when fitting on smaller scales.

Clearly, our partial statistical analysis does not yield a
rigorous bound on the magnitude of $B$. However,
the results in Fig.~\ref{fig:chiminall} can be used to get a rough 
estimate for the allowed range of the $B$ values. 
From the discussion at the start of this section, a reasonable 
bound can be based on taking $\chi^2(B)/\chi^2(0)\lesssim 1.25$.
In Fig.~\ref{fig:chiminall}, the maximum value of $B$
corresponding to this criterion would be $B_{max} \sim 0.1$.
Moreover, we saw that the actual
minimum value of $\chi^2(B)$ in the right panel of Fig.~\ref{fig:chiminall}
was at $B\simeq 0.025$, not zero, which indicates that, almost
certainly, the available
data cannot exclude values of $B\lesssim 0.025$. As discussed in \S~\ref{model},
this does not simply translate into a constraint on the value of the string
tension, $G\mu$. Instead, from Eq.~(\ref{bdep}), $B \propto (G\mu^/\lambda)^2$.

\begin{figure}[!]
\scalebox{0.420}{\includegraphics{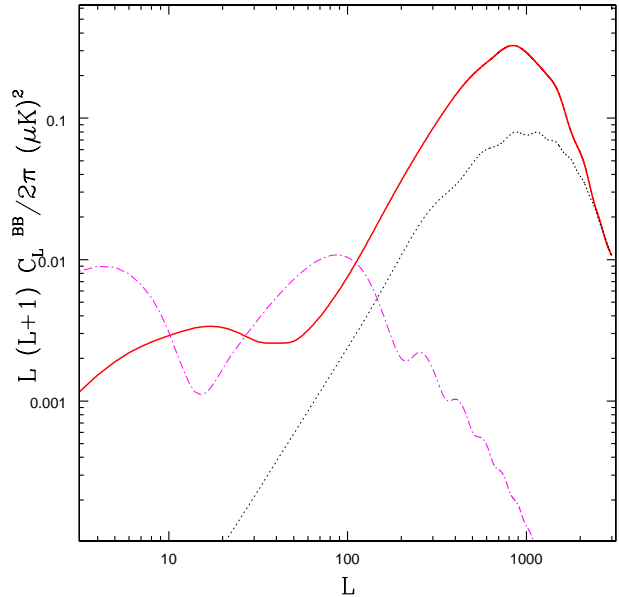}}
\caption{\label{fig:clbb}The $B$-type polarization spectrum $C_l^{BB}$
due to cosmic strings
as predicted by our string model with the fit parameter $B$ set to $0.1$ is
plotted as a solid red line. The expected  $C_l^{BB}$ spectra for
E to B lensing (dotted black line) and from primordial gravitational waves
assuming a tensor-to-scalar ratio of $r=0.1$ (dash-dotted magenta line) are
plotted for comparison.}
\end{figure}

Because cosmic strings produce a sizeable vector perturbation, in addition to
the $E$-polarization, they can also induce the $B$-type polarization of the CMBR.
In Fig.~\ref{fig:clbb} we plot the $B$-polarization angular spectrum,
$C_l^{BB}$, as predicted by our model with $B=0.1$.
It appears that, even if cosmic strings were contributing only a few percent to the
CMB temperature anisotropy, they could contribute a 
B-mode signal on scales $l\sim 800$ that would be in excess of the expected 
B-mode produced by lensing of the E-mode by galaxies \cite{lensing}.
Such a detection would be an important test for the existence of cosmic strings.

\section{Discussion}
\label{discussion}

Cosmic strings appear to be a likely by-product of inflation in
superstring cosmology. Although these cosmological models are
only just beginning to be developed in sufficient detail for
comparisons with data to be possible, most models seem to feature
a string scale comparable to or slightly smaller than the GUT
scale, and cosmic string tensions $G\mu\sim 10^{-10}-10^{-6}$.

These expectations motivate studying models in which CMB
fluctuations and large scale structure are the consequence
of both growing modes associated with perturbations generated
from quantum fluctuations during inflation, and active perturbations
associated with the gravitational effects of a network of cosmic
strings. Here, we have presented an attempt
at such an analysis,
based on a rough comparison of theoretical models with both
adiabatic perturbations and cosmic strings with WMAP TT and TE power
spectra as well as the 2dFGRS results on galaxy correlation functions.
Our results need to be improved in various ways, both by
allowing numerous cosmological parameters we held fixed to vary,
and by computing likelihood functions, rather than using a rough
criterion based on a reduced $\chi^2$ statistic. Nevertheless,
they already indicate that although the data currently available
generally favor models without cosmic strings, they may not exclude
nonzero cosmic string tension and density definitively, provided the growth
of fluctuations in the Universe is not dominated by cosmic strings.
Rather conservatively, we have argued that values of $B\lesssim
0.1$ are not excluded, although this result needs to be put on
firmer footing statistically; we are quite confident that
$B\lesssim 0.025$ cannot be excluded as yet.
Since $\sqrt{B} \simeq (0.25/\lambda) (G\mu/1.1\times 10^{-6})$,
these values of $B$ would correspond to
\be
G\mu\lesssim {3.5 \times 10^{-7} {\lambda \over 0.25} \sqrt{B\over 0.1}}~.
\ee
We have also found the values of $n_s-1$ that minimze the reduced
$\chi^2$ statistic for different
values of $B$. Conservatively, for the range $B\lesssim 0.1$,
the various different comparisons with data in the two panels of
Fig.~\ref{fig:chiminall} are consistent with $n_s-1\simeq -(0.04-0.01)$.
Comparing with the predictions of various brane inflation models via
Eqs. (\ref{nsdnscoulomb}), (\ref{nsdnspl}) and (\ref{nstwo}), we note
that the values we find are in agreement and certainly not in
conflict with any of the models.
However, we note that the Coulombic inflation models
generally predict $\vert n_s-1\vert\lesssim-0.03$, at the low end of
the range of values we infer from comparing with the data, while
the powerlaw
models allow larger $\vert n_s-1\vert$; Eq. (\ref{nsdnspl}) implies
$\vert n_s-1\vert\gtrsim 0.03$ for any value of $\sigma$, and Eq. (\ref{nstwo})
implies a constant powerlaw index whose value depends on parameters of
the brane inflation model.
If $n_s(k)$ can be pinned
down with greater precision, it may become possible to discriminate among
different brane inflation models.
We caution, though, since we have not varied the background cosmological model in
obtaining these bounds, we cannot rule out that a more complete analysis
would allow still larger values of $G\mu$, with other values of
$n_s-1$. In a future publication we
intend to present results of a more comprehensive study which would include
varying all cosmological parameters as well as the relevant parameters of the
string model, including their specific predictions for $n_s(k)$, and a better
justified statistical analysis based on codes made available by the WMAP
team for computing likelihood functions for any model for the production
of temperature pertubations \cite{wmap_verde}. (Some modifications will
be needed to account for the nongaussianity of the cosmic string perturbations.)
In addition to the WMAP and the 2dFGRS data, the analysis will
also include the latest results from the Sload Digital Sky Survey (SDSS) \cite{sloan}.

The key to our analysis is the idea that the Universe is a patchwork quilt,
with a little bit of cosmic strings thrown in to complicate the models.
A smoking gun for the existence of cosmic strings could be the detection
of B-mode polarization on smaller scales at an amplitude considerably
larger than is predicted in inflation models without cosmic strings.
Another potentially distinguishing signature of cosmic strings could be
the detection of cosmological non-Gaussianity.
Tests of the WMAP data \cite{wmap_komatsu} have so far been limited to
constraining the type of non-Gaussianity expected from inflationary models.
Some of the most commonly used tests of non-Gaussianity, such as the
bispectrum test, had actually been shown to be insensitive to possible contributions
from cosmic strings \cite{gangui}. Specially tailored tests are likely to
be needed to detect string induced non-Gaussianity \cite{proty}.
Cosmic strings may also be detected from the observation of identical galaxy pairs
in close proximity on the sky \cite{galaxypairs}. Gravitational radiation
from kinks in cosmic strings may also be detectable down to exceptionally
small values of $G\mu$ \cite{vil-dam}.
Pulsar timing is likely to push bounds on
the density in long wavelength gravitational radiation backgrounds
\cite{kaspi}
down by an order of magnitude or so, corresponding to a factor of
three or so in $G\mu/\lambda^2$ \cite{backer}, but a substantial
contribution to the background from supermassive black hole
binaries \cite{jaffe} may frustrate our
ability to use these observations to constrain the cosmic string
tension much further.

\acknowledgments
We thank Richard Battye, Nick Jones, Will Kinney, Horace Stoica,
Alex Vilenkin, and Neal Weiner for discussions, and Tom Loredo and Saul Teukolsky
for critical comments. This research is partially supported by
NSF (S.-H.H.T.), NASA (I.W.) and PPARC (L.P.). M.W is supported by the
NSF Graduate Fellowship. I.W. acknowledges the hospitality of KITP,
which is supported by NSF grant PHY99-07949, where part of this research was carried out.

\end{document}